\documentstyle[epsfig,nicV]{article}
\begin{document}
%
%
\heading{%
Experimental determination of the $^{17}$O(n$_{th}$,$\alpha $)$^{14}$C
reaction cross section
%
}
\par\medskip\noindent
%
\author{Jan Wagemans$^{1}$, Cyriel Wagemans$^{2}$, Ronald Bieber$^{1}$ and
Peter Geltenbort$^{3}$
}
\address{
EC, Joint Research Centre, Institute for Reference Materials and
Measurements, Retieseweg, B-2440 Geel, Belgium}
\address{
Department of Subatomic and Radiation Physics, RUG,
Proeftuinstraat 86, B-9000 Gent, Belgium}
\address{
Institut Laue-Langevin, Boite Postale 156, F-38042
Grenoble, France}

\begin{abstract}
The $^{17}$O(n$_{th}$,$\alpha $)$^{14}$C reaction cross section was
determined at the high flux reactor of the ILL in Grenoble relative to the
known $^{14}$N(n$_{th}$,p)$^{14}$C cross section. The $^{17}$O(n$_{th}$,$%
\alpha $)$^{14}$C measurements were performed with several highly enriched
oxygen gas samples and the flux calibration was done with $^{14}$N$_{2}$
from the air. This resulted in a precise value of (244 $\pm $ 7) mb for the $%
^{17}$O(n$_{th}$,$\alpha $)$^{14}$C cross section.
\end{abstract}
\section{Introduction}
Inhomogeneous big bang models are able to reproduce the primordial light
element abundances, but moreover lead to the production of potentially
observable nuclides heavier than C \cite{3}. In addition, these models could
solve conflicts that arise in the standard big bang model \cite{4}. At
present it is thought that $^{14}$C may act as a bottleneck in the
nucleosynthesis path to heavier elements. In this respect, Applegate {\it et
al}. \cite{1} and Thielemann {\it et al}. \cite{10} demonstrated the
dominant influence of the $^{17}$O(n,$\alpha $)$^{14}$C reaction on the
production of elements heavier than A = 17, which will be strongly hindered
by this reaction since it recycles the mass flow back to $^{14}$C.

The $^{17}$O(n,$\alpha $)$^{14}$C reaction could also help to explain
anomalies in $^{18}$O/$^{16}$O and $^{17}$O/$^{16}$O ratios found in
presolar grains \cite{8}. These grains are stellar condensates and their
isotopic compositions reflect both the initial compositions of stars and the
changes that arise as a result of nucleosynthesis and stellar evolution.

So far, only two direct measurements of the $^{17}$O(n,$\alpha $)$^{14}$C
reaction cross section are available at relevant neutron energies \cite{5},
\cite{9},which both completely rely on a value of 235 mb for the $^{17}$O(n$%
_{th}$,$\alpha $)$^{14}$C reaction cross section. This value definitely
needs to be verified since it is based on only two old experimental cross section
determinations: (560 $\pm $\ 130) mb \cite{6} and (235 $\pm $\ 5) mb \cite{2}.
\section{Experimental method}
\subsection{Experimental setup}
The measurements were performed at the end of the thermal neutron guide H22D
of the high flux reactor of the ILL in Grenoble (France). The thermal flux
at the sample position reached a value of about 5 $\times $10$^{8}$ n/cm$%
^{2} $s, the ratio of thermal neutrons to epithermal and fast neutrons was 10%
$^{6} $ and the $\gamma $ ray flux from the reactor was reduced by a factor
of 10$^{6}$. The flux determination was done based on the known $^{14}$N(n$%
_{th}$,p)$^{14}$C cross section, which permits to perform these measurements
with gaseous samples.

The reaction studied and the flux calibration should be performed in exactly
the same detection geometry and in both cases the number of atoms should be
well known. To do this with gaseous samples, we rely on the basic
principle that 1 mole (6.022 $\times $10$^{23}$ atoms) of any gas at a
pressure of 1 atm. fills a volume of 22.414 liter. When a volume of oxygen
gas and afterwards the same volume of air (containing 77.8 \% $^{14}$N) is
injected in the vacuum chamber, the gas resp. the air will be distributed
homogeneously in the chamber. So the $^{17}$O resp. $^{14}$N density in the
beam profile will be simply proportional to the resp. enrichment factors of
both gasses.

Since the $^{17}$O(n$_{th}$,$\alpha $) and $^{14}$N(n$_{th}$,p) particles
have low energies (1.4 MeV and 0.6 MeV resp.), special care had to be given
to the detector choice. We used two (collimated) fully depleted surface barrier
detectors (thicknesses 14.8 $\mu $m and 21.3 $\mu $m, resolution 55 keV and
31 keV) which have an excellent signal-to-noise ratio in the region of
interest. Their energy calibration was done by means of the $^{14}$N(n$_{th}$%
,p)$^{14}$C, $^{10}$B(n$_{th}$,$\alpha $)$^{7}$Li and $^{6}$Li(n$_{th}$,$%
\alpha $)t reactions. Both detectors were mounted in a
vacuum chamber, parallel with and outside of the neutron beam.
\subsection{Measurements}
The measuring sequence consists of three gas injections. Before each
injection a vacuum of 10$^{-3}$ torr (see figure 1) is
established in the system. First, the volume of air between the glass seal S
and valve 2 is injected in the vacuum chamber and the $^{14}$N(n$_{th}$,p)
particles counted. Then, the glass seal is broken by means of the steel rod
\begin{center}
\begin{figure}[hbt]
{\vbox{\epsfig{file=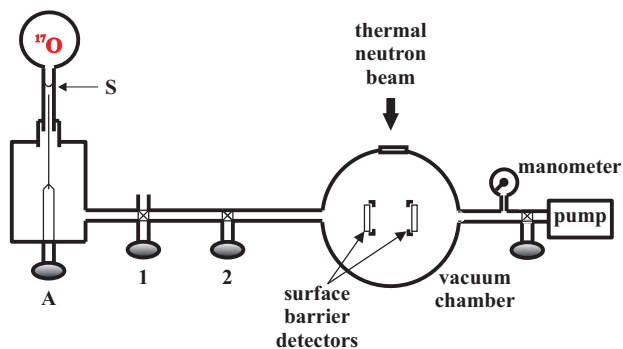,width=9.2cm}}}
\caption{
Schematic picture of the setup (not on scale!) showing the
gas injection system.
}
\label{figure1}
\end{figure}
\end{center}
A, the oxygen is injected and the $^{17}$O(n$_{th}$,$\alpha $) particles are
counted. Finally, the second part of the flux measurement is executed: air
contained in the volume from valve 2 and the bottle
(since the glass seal is broken!) is injected in the chamber and again the $%
^{14}$N(n$_{th}$,p) particles are counted. When subtracting the first
measurement from the third one, the $^{14}$N(n$_{th}$,p) counting rate
corresponding to the air in the {\it sealed part} of the bottle is obtained.
This sequence was repeated with several bottles containing 100 ml of oxygen
gas (1 atm.) with $^{17}$O enrichments of 58.2 at.\% and 85.5 at.\%. Typical
results are shown in figure 2. On the left side we see the pulse height
spectrum for the $^{17}$O(n$_{th}$,$\alpha $) reaction obtained 15 hours
after the injection of 85.5 at.\% enriched $^{17}$O gas. The right side of the
figure shows the $^{14}$N(n$_{th}$,p) flux measurement. Both spectra are
corrected for (small) background contributions.
\begin{center}
\begin{figure}[htb]
{\vbox{\epsfig{file=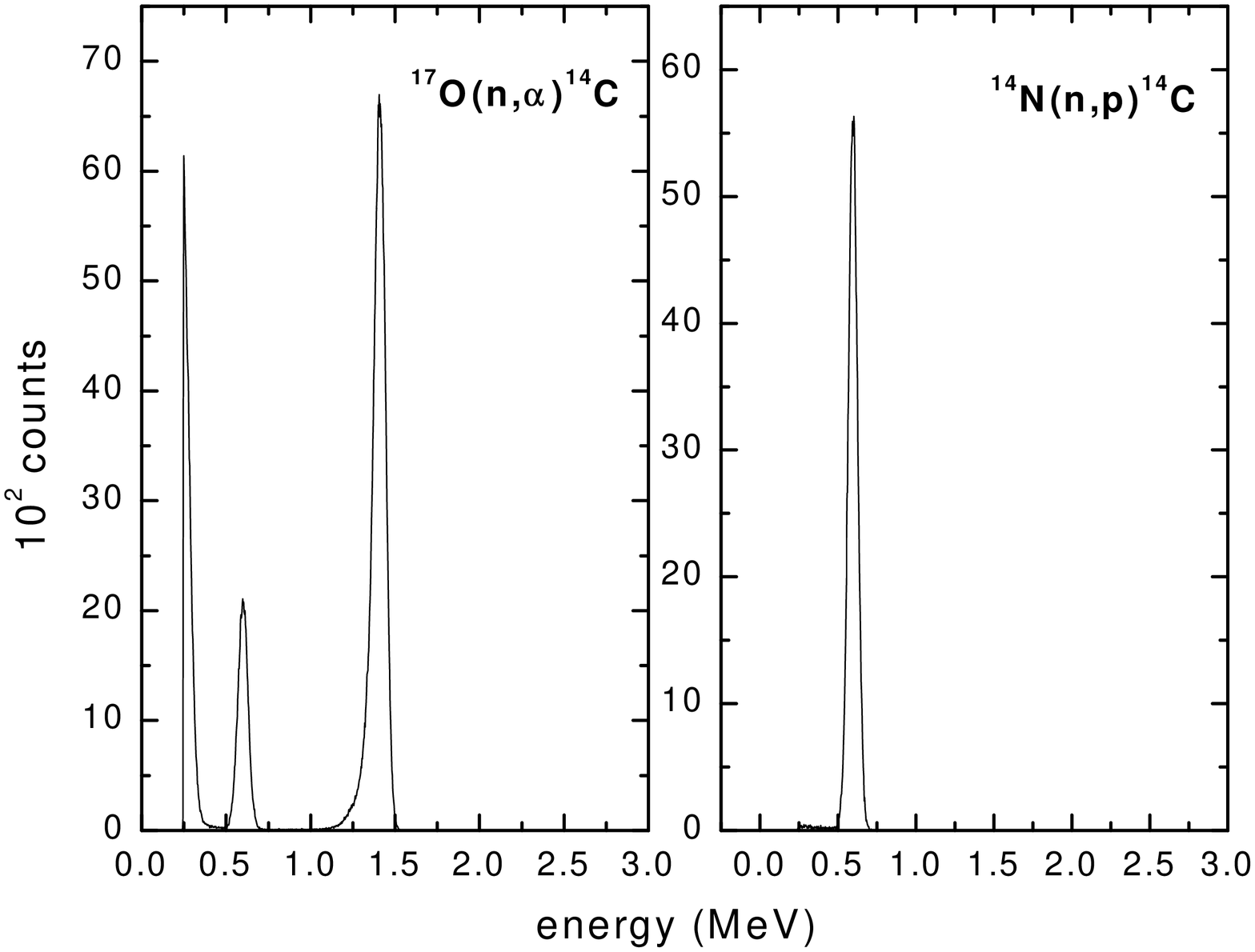,height=4.7cm,width=8.5cm}}}
\caption{
Energy distribution for the $^{17}$O(n$_{th}$,$\alpha $)$^{14}$C reaction 
and for the flux determination $^{14}$N(n$_{th}$,p)$^{14}$C.
}
\label{figure2}
\end{figure}
\end{center}
\section{Results and discussion}

The $^{17}$O(n$_{th}$,$\alpha $)$^{14}$C reaction cross section is
determined relative to the $^{14}$N(n$_{th}$,p)$^{14}$C reaction, using the
relation 
\begin{equation}
\sigma _{\alpha }(^{17}O)=\sigma _{p}(^{14}N)\frac{Y_{\alpha }(^{17}O)}{%
Y_{p}(^{14}N)}\frac{N(^{14}N)}{N(^{17}O)}  \label{one}
\end{equation}
in which $\sigma $ is the thermal cross section value, Y the counting rate
after background subtraction and N the number of atoms per cm$^{3}$.

As mentioned earlier, the N($^{14}$N)/N($^{17}$O) \ density ratio will be
equal to the isotopic enrichment ratio I($^{14}$N)/I($^{17}$O) of the gasses
used. The I($^{14}$N) value still needs to be multiplied by a factor f(p,h)
to correct for the meteorological conditions during the measurements,
because the filling of the glass bottle with atmospheric air is sensitive to
the atmospheric pressure (p) and the humidity of the air (h) at that moment. For the 
$^{14}$N(n$_{th}$,p) cross section, we adopted the evaluated value of (1.83 $%
\pm $ 0.03) b, so relation (\ref{one}) becomes: 
\begin{equation}
\sigma _{\alpha }(^{17}O)=1.83\frac{Y_{\alpha }(^{17}O)}{Y_{p}(^{14}N)}\frac{
I(^{14}N)}{I(^{17}O)} f(p,h)\,b  
\label{two}
\end{equation}

Two experiments were performed with oxygen gas with an enrichment of 58.2
at.\% in $^{17}$O and with only one detector mounted and three with gas
enriched to 85.5 at.\% and two detectors mounted. Hence eight values for the 
$^{17}$O(n$_{th}$,$\alpha $) cross section could be calculated using
relation (\ref{two}). The corresponding results are displayed in figure 3 
, which demonstrates the reproducibility of our method.
\begin{center}
\begin{figure}[htb]
{\vbox{\epsfig{file=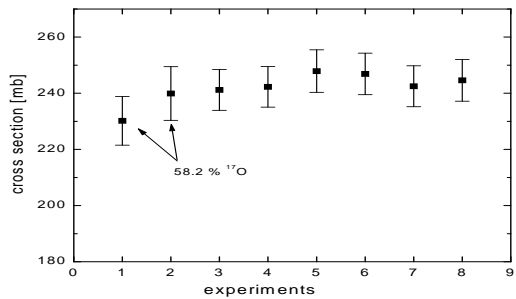,height=4.7cm,width=8.5cm}}}
\caption{
The $^{17}$O(n$_{th}$,$\alpha $)$^{14}$C cross section determined
in the eight different runs.
}
\label{pippo}
\end{figure}
\end{center}
The uncertainty on the results is composed of three \ contributions: (i) a
statistical error varying between 0.4 \% and 1.4 \%; (ii) the uncertainty on
the meteorological correction factor, for which a conservative value of 1 \%
was adopted; (iii) a systematic normalisation uncertainty of 1.6 \% due to
the uncertainty on the $^{14}$N(n$_{th}$,p) cross section. A weighted
average of the eight runs was calculated using the statistical errors,
resulting in a value of (244 $\pm $ 7) mb for the $^{17}$O(n$_{th}$,$\alpha $%
) cross section.

This result agrees within the experimental uncertainties with the value of
Hanna {\it et al}.\cite{2}. They used CO$_{2}$ gas with an isotopic
enrichment in $^{17}$O of $\approx $ 0.3 \% and 2 \% and determined the $%
^{17}$O(n$_{th}$,$\alpha $) cross section relative to the activation cross
section of gold. They performed six measurements resulting in an average
value of (235 $\pm $ 5) mb.
\section{Conclusions}
The present result combined with the value of Hanna et al. \cite{2}
converges to a value of about 240 mb for the $^{17}$O(n$_{th}$,$\alpha $)
cross section. This confirms the correctness of the normalisation of the $%
^{17}$O(n,$\alpha $) measurements at higher neutron energies \cite{5},\cite
{9}, which relied on a thermal value of 235 mb. This in turn increases the
confidence in the astrophysical reaction rates calculated from these higher
energy data.

\acknowledgements{This research was sponsored by Nato Collaborative Research
Grant 960455 and by the Fund for Scientific Research Flanders }

\vfill
\end{document}